# Interaction-driven (quasi-) insulating ground states of gapped electron-doped bilayer graphene

Anna M. Seiler[1], Martin Statz[1], Isabell Weimer[1], Nils Jacobsen[1], Kenji Watanabe[2], Takashi Taniguchi[3], Zhiyu Dong[4], Leonid S. Levitov[4], R. Thomas Weitz[1]*

[1]1st Physical Institute, Faculty of Physics, University of Göttingen, Friedrich-Hund-Platz 1, Göttingen 37077, Germany

[2]Research Center for Functional Materials, National Institute for Materials Science, 1-1 Namiki, Tsukuba 305-0044, Japan

[3]International Center for Materials Nanoarchitectonics, National Institute for Materials Science, Tsukuba, Japan

[4]Department of Physics, Massachusetts Institute of Technology, Cambridge, Massachusetts 02139, USA

*Corresponding author. Email: thomas.weitz@uni-goettingen.de

**Bernal bilayer graphene has recently been discovered to exhibit a wide range of unique ordered phases resulting from interaction-driven effects and encompassing spin and valley magnetism, correlated insulators, correlated metals, and superconductivity. This letter reports on a novel family of correlated phases characterized by spin and valley ordering, observed in electron-doped bilayer graphene. The novel correlated phases demonstrate an intriguing non-linear current-bias behavior at ultralow currents that is sensitive to the onset of the phases and is accompanied by an insulating temperature dependence, providing strong evidence for the presence of unconventional charge carrying degrees of freedom originating from ordering. These characteristics cannot be solely attributed to any of the previously reported phases, and are qualitatively different from the behavior seen previously on the hole-doped side. Instead, our observations align with the presence of charge- or spin-density-waves state that open a gap on a portion of the Fermi surface or fully gapped Wigner crystals. The resulting new phases, quasi-insulators in which part of the Fermi surface remains intact or valley-polarized and valley-unpolarized Wigner crystals, coexist with previously known Stoner phases, resulting in an exceptionally intricate phase diagram.**

Graphene flat-band systems, moiré and non-moiré, provide a rich platform in which a wide variety of correlated electronic states can be realized and explored[1–13]. After initial observations of exchange driven phases in suspended Bernal bilayer graphene at charge neutrality[5,6,10], a family of exotic interaction-driven orders has been found in magic-angle twisted bilayer graphene (TBG), where flat electronic bands are created by superposing two layers of graphene with a relative twist angle. In TBG, Mott insulating and superconducting phases have been found to compete at half filling of either electron or hole Moiré bands[7,8]. However, it quickly became evident that TBG does not possess an exclusive monopoly on hosting interesting correlated states in the vicinity of van-Hove-singularities. These states have also been found in the naturally occurring Bernal bilayer graphene (BLG), where strongly correlated behavior arises at large electric displacement fields ($D$), assisted by field-induced flat bands and a divergence in the density of states (DOS)[1–4,14] at Lifshitz transitions. Close to the band edge interactions between charge carriers are large enough to satisfy the Stoner criterion[15] and carriers become spin or valley polarized or inter-valley coherent, as can be deduced from varying Landau level (LL) spacings at finite magnetic fields $B$[1–3]. Previous studies on BLG focused on the hole-doped regime where several exotic orders beyond Stoner magnetism such as correlated and superconducting[1,3,4] phases have been observed. These states have often been attributed to the presence of annular Fermi surfaces or pocket physics, as found on the hole side in BLG[1,3] and in ABC trilayer graphene[11,12]. Yet, at this point it is far from clear whether the correlated phases are driven primarily by the band dispersion or by some exotic interaction effects[16].

To shed light on this question, here we report on the previously unexplored electron-doped regime of BLG. The low-energy dispersion on the electron side is markedly different compared to the hole side. For example, the trigonal warping induced rotational symmetry breaking is significantly less pronounced at the conduction band minimum (Fig 1a,b). In fact, at large $D$, the bands are flatter and the DOS-peak due to the still remaining Lifshitz transition is significantly stronger than in the valence band (Fig. 1c, Extended Data Fig. 1). Nevertheless, so far, only metallic Stoner phases have been observed in electron-doped Bernal-stacked bilayer graphene[1,2]. Surprisingly, we find a new interaction-driven behavior that is clearly beyond Stoner physics. Specifically, complex correlated phases appear close to the band edge at large $D$ and low magnetic fields ($B$) that cannot be explained solely by Stoner ferromagnetism. As discussed below, the isospin polarized orders in this doping regime show an insulating temperature dependence and a striking nonlinear current-bias dependence. These behaviors are distinct from the previously reported results[2], indicating an unexplored ground state of the electron-doped bilayer graphene. There are several theoretical pictures in recent literature that account for these observations, such as formation of a spin or charge density wave order that gaps out part of the Fermi surface[17] or a Wigner crystal, potentially coexisting with mobile carriers[1].

The bilayer graphene flakes investigated for this study are encapsulated in hexagonal boron nitride (hBN) and equipped with graphite top and bottom gates, which allows to continuously tune the displacement field $D$ as well as the charge carrier density $n$ (see methods). Measurements shown in the main text were conducted in a device with two-terminal graphite contacts (Fig. 1d), which is why all conductance data except the data presented in Extended Data Fig. 2 is given in arbitrary units (a.u., see methods for details). Measurements conducted in a second device with four-terminal edge contacts, which shows qualitatively similar behaviour to the device discussed in the main manuscript, are shown in Extended Data Fig. 3. All measurements, unless stated otherwise, were conducted in a dilution refrigerator at a base temperature $T$ of 10 mK.

Figure 1e shows the measured two-terminal conductance $G$ as a function of $n$ and $D$ for electron doping and negative $D$. At charge neutrality, an electric-field induced band gap opens up resulting in a well-known conductance minimum[5]. Outside the gap, $G$ increases monotonically as a function of $n$. At large $D$, where interactions are expected to be strong due to a peak in the DOS (Fig. 1c), two steps emerge, separating regions of similar conductance that we tentatively label with **svi** (spin and valley polarized insulator), **si** (spin polarized insulator), and **m** (metal). In the following we will discuss in further detail why we chose these labels. The regions of similar conductance can be differentiated best in the normalized derivative of $G$ with respect to $n$ (|d$G$/d$n$|), where the steps in between them appear as peaks (Fig. 1f, see Extended Fig. 3 for a corresponding 4-terminal measurement). Parts of these features have been observed previously in capacitance measurements without magnetic fields ($B$ = 0 T) and at finite out-of-plane magnetic fields $B_\perp$ > 0.6 T [Ref. [2]]. In Ref. [2], one of the peaks is suggested to mark the transition between a Stoner quarter and a full metal, corresponding to a region with one-fold and another region with four-fold degenerate LL. We can reproduce these results (both the results at B = 0 T and at $B_\perp$ > 0.6 T) and find regions with one-fold and four-fold degenerate quantum Hall states at $B_\perp$ > 0.3 T (Extended Data Fig. 4, 5).

At present, however, the nature of the phases at $B$ = 0 T is unclear since their nature had previously only been extrapolated from the high-magnetic field behavior (at $B_\perp$ > 0.6 T)[2]. This means, while previous studies focused on Stoner ferromagnetism in bilayer[1–3] and trilayer graphene[11] at finite magnetic fields, the true $B$ = 0 T ground state of the system remains unknown. In fact, at magnetic fields $B_\perp$ < 0.3 T, we were not able to observe quantum Hall states in the density regime of the **svi** and **si** phases (Fig. 2a-c), while quantum Hall states are well visible even below 200 mT in the **m** phase (Fig. 2c, Extended Data Fig. 5). This observation clearly indicates that the phases studied at elevated magnetic field likely are distinct in nature from the truly interacting $B$ = 0 T ground state.

To analyze the nature of these phases, we first study the system around $B$ = 0 T as function of $D$. At constant $D$ = -0.6 V nm$^{-1}$, all three phases are well resolved at $B_\perp$ = 0 T . With slightly increasing $B_\perp$ (Fig. 2a), the mutual phase boundary of the **svi** phase and the **si** phase shifts towards larger densities i.e., the **svi** phase becomes stable in a larger density region. Above $B_\perp \approx 200 - 300$ mT, the phase boundary between the **svi** and **si** phases becomes less pronounced or vanishes. Remarkably, the **si** phase is even less stable against $B_\perp$ at large $D$-fields (Fig. 2b,c) while at $D$-fields below $|D| = 0.5$ V nm$^{-1}$, small $B_\perp$ can even stabilize this phase (Fig. 1f, Fig. 2c). Since in BLG the energy scale given by the coupling between the out-of-plane magnetic field and the valley degree of freedom via the orbital momentum is larger compared to other magnetic field induced splittings (e.g. Zeeman)[10], we conclude that the **svi** and the **si** phases must have different valley orderings. In the simplest case, the **svi** phase is then valley polarized due to exchange-driven Stoner physics, while the **si** and **m** phases are valley unpolarized[2,11]. Another possibility is inter-valley coherence in the svi phase[1]. Notably, the phase boundary of the **si** phase and the **m** phase shifts also slightly with $B_\perp$, likely resulting from different spin orderings in these phases that are discussed below.

Next, we investigate the phase transitions as a function of the in-plane magnetic field $B_{||}$ that primarily couples to the spin degree of freedom via the Zeeman effect (Fig. 2d). Whereas increasing $B_{||}$ does not change the phase boundary between the **svi** and the **si** phase (Fig. 2d), the phase boundary between the **si** and the **m** phase moves towards higher densities marking a phase transition with different spin polarization[11]. Since the **si** phase becomes broader with increasing $B_{||}$, we assume the **si** phase to be spin polarized, while we suggest the **m** phase to be spin degenerate. The **svi** phase is likely spin polarized as well since the phase boundary between the **si** and **svi** phase does not move in density with increasing $B_{||}$. At $|D| < 0.6$ V/nm, where the **svi** and **si** phases do not exist at zero magnetic field, $B_{||}$ stabilizes the spin polarized **si** phase (Fig. 2e,f).

The assignment of spin and valley polarization is consistent with the LL degeneracies discussed above[2,11] and also with magnetic hysteresis measurements shown in Extended Data Fig. 5. While the spin polarized **svi** and **si** phases show an in-plane magnetic field hysteresis due to spin polarization (Extended Data Fig. 5a), the spin and valley polarized **svi** phase exhibits also an out-of-plane magnetic field hysteresis resulting likely from orbital magnetism as there is no out-of-plane magnetic field hysteresis in the **si** phase (Extended Data Fig. 5b). Interestingly, in the **svi** phase, the in-plane magnetic field hysteresis induced by spin magnetism is despite weak spin-orbit coupling in graphene much larger than the out-of-plane magnetic hysteresis induced by orbital magnetism, which might be explained by an inter-valley coherent component of this phase[18]. For example, at $D = -0.6$ Vnm$^{-1}$, the hysteretic behavior in the in-plane magnetic field ends at $B_{||} = \pm 0.6$ T while in an out-of-plane magnetic field it is only present until $B_{\perp} = \pm 0.03$ T.

To conclude our initial observation, three phases with spin and valley ordering have been identified in electron-doped BLG close to the band edge were electron-electron interactions are strong: the **svi** phase is likely spin and valley polarized, the **si** phase is likely spin polarized and the **m** phase is spin and valley unpolarized.

While we revealed different regions of spin and valley polarization, our measurements up to now cannot identify if the observed phases are conventional Stoner phases in which e.g., the absence of LL at low magnetic fields stem from flat electronic bands with concomitant large effective masses and small LL spacings, or if the ground state of the system is of a more exotic nature. This is why we have performed detailed temperature and bias-dependent measurements (Fig. 3 and 4). Remarkably, close to the base temperature of our cryostat of 10 mK, we observe an increasing conductance with increasing temperature (insulating temperature dependence) in the **svi** and **si** phases which suggests these phases to be different from normal Stoner metals, where a metallic temperature dependence would be expected. This insulating temperature dependence is likely not connected to contact effects since it is only present within the **svi** and **si** phases but not at the phase boundaries which can be seen best in Fig. 3b (see also below).

The **svi** phase displays an insulating temperature dependence until $T \gtrsim 4$ K (Fig. 3, Extended Data Fig. 6). Since it is closest to the band edge where the DOS diverges (Fig. 1c), correlation effects are expected to be the most prominent in this phase. The **si** phase displays an insulating temperature dependence in a comparably lower temperature range (critical temperature $T_{crit} < 100$ mK at $D = -0.6$ V nm$^{-1}$, Fig. 3a), but $T_{crit}$ can be increased by applying a finite $B_{||}$ (e.g. $T_{crit} \approx 2.5$ K at $B_{||} = 2$ T, Fig. 3b, Extended Data Fig. 7). The **m** phase only shows insulating behavior at $T_{crit} < 100$ mK in the vicinity of the phase boundary to the **si** phase but acts weakly metallic at large densities (even at the lowest $T$) as expected in the non-interacting regime of bilayer graphene (Fig. 4, Extended Data Fig. 8). The observation of insulating temperature dependence in the **svi** and **si** phases aligns well with the measured bias current ($I$) dependence that shows strong nonlinearities and is indicative of a comparably larger gap in the **svi** phase and a smaller one in the **si** phase where it vanishes at small applied bias currents of approximately 10 nA at $D = -0.8$ V nm$^{-1}$ (Fig.

4a,b, Extended Data Fig. 9,10). It resembles the bias current dependence of correlated phases in hole-doped bilayer graphene where similar behavior was suggested to be consistent with the formation of charge or spin density waves (CDW or SDW)[3] or Wigner crystals (WC)[1] in which Coulomb interaction breaks translation symmetry and strong electron-electron interactions lead to a collective ordering of electrons. In Ref. [1], the correlated insulating phases at hole-doping (phase II and phase III in Ref. [1]) appear when the Fermi surface is annular and, thus, in a different density regime than the Stoner phases observed at hole-doping[1–3]. However, in the electron-doped regime discussed in this work, the spin and valley ordered phases themselves become insulating.

One possibility for this behavior is that the Stoner ferromagnetic states transition into Wigner crystal states that have different spin and valley orderings or, alternatively, into charge density wave (CDW) or spin density wave (SDW) states. Indeed, for a 2D system in a Stoner regime that is introduced above, theory predicts a CDW or SDW instability, arising because the momentum dependence of particle-hole susceptibility for 2D systems is in general non-monotonic - the susceptibility at a finite momentum $q$ can exceed that at $q = 0$. The resulting CDW or SDW instability with a wavelength corresponding to the most unstable harmonic $q$, will modulate the Stoner-polarized Fermi sea in electron density or spin density[17].

The hallmark of such electronic ordered states, CDWs, SDWs, or WCs, is a highly nonlinear transport arising due to CDW, SDW, or WC sliding in the presence of an applied bias current. Indeed, we observe a conductance in the **svi** and **si** phase that peaks at a finite current (see Figure 4a). One possible explanation for this intriguing behavior involves the pinning of CDW/SDW/WC sliding at low currents and its subsequent depinning at higher currents[19]. Essentially, the conductivity experiences a sharp increase beyond a critical current that induces the depinning of the CDW/SDW/WC and initiates the sliding motion. However, as the current continues to rise, the conductivity begins to decline. This can be potentially attributed to the Doppler shift of the particle-hole excitation dispersion caused by the sliding CDW/SDW/WC, eventually leading to the closure of the CDW/SDW/WC gap at high currents[20,21]. Consequently, the conductivity at high currents remains slightly above that at low currents.

This interpretation is consistent when studying the evolution of nonlinear conductance under varying carrier density (Fig. 4a and b, Extended Data Fig. 9, 10). The peak current decreases continuously to zero when approaching the phase transition into **svi** phase and **m** phase. Following our interpretation above that the peak corresponds to the CDW/SDW/WC gap, this observation implies a gap closing at the phase boundary (Extended Data Fig. 11).

Next, we discuss how the 2D CDW/SDW/WC picture explains the measured conductance temperature dependence described above. WCs are fully gapped and consequently show an insulating temperature dependence[1]. Also a CDW/SDW in 2D opens up gaps in small segments near the Fermi surface regions where nesting occurs, whereas the rest of Fermi surface remains gapless. The coexistence of gapped and gapless segments leads to two effects:

(i) As temperature rises, the CDW/SDW gap decreases. In that, the length of the gapped segments also decreases, whereas the gapless segments increase. As a result, the effective carrier density in this system increases. This effect leads to conductivity monotonically increasing with temperature.

(ii) However, as in ordinary metals, the conductivity at a fixed carrier density should decrease as temperature increases. This effect tends to give a conductivity monotonically decreasing with temperature.

These two effects compete with each other, leading to two possible outcomes. If the effect (i) is stronger, the system will mimic an insulator's behavior where a conductivity monotonically increases with temperature. Yet, unlike a true insulator, this increase of conductivity will be weak. We call such a system a quasi-insulator . This prediction is indeed in line with our measurement in **svi** and **si** phases (Fig. 4a,b, resistance values in these phases are shown in Extended Data Fig. 2).

Surprisingly, also the **m** phase exhibits an insulating temperature dependence and a non-linear bias current dependence in the vicinity of the phase boundary between the **si** and **m** phase (but not deep within the metallic phase, Extended Data Fig. 8, 11) which we explain by fluctuations of the collected electronic mode near the second order phase transition.

Concluding, our experimental data is in line with the picture of fully gapped WCs or CDWs/SDWs that are coexisting with the Stoner phases and gap out parts of the Fermi surface. To unambiguously determine the kind of ordering, further measurements, e.g. using scanning tunneling microscopy[22] or angle-resolved photoemission spectroscopy[23], would be needed.

In summary, electron-doped bilayer graphene exhibits an intriguing phase diagram at small magnetic fields characterized by three phases that exhibit different spin and valley ordering (a summary of the experimental findings is given in Extended Data Table 1). We find a (quasi-)insulating temperature dependence (conductance growing with temperature) and a nonlinear current bias dependence in two phases and consequently name them as spin and valley polarized insulators (**svi**) and spin polarized insulators (**si**). The high-density metal phase (**m**), on the other hand, is consistent with a non-interacting phase. In contrast to previously studied hole-doped bilayer graphene, the phases are only stable at very low $n$ and do not seem to be related to changes in the Fermi surface topology within the band but rather to flat electronic bands and divergent DOS at the band edge. Here, the DOS seems to be large enough to facilitate the formation of an exotic ordering of electrons such as in a CDW/SDW or WC. Similar orderings might also appear in electron-doped rhombohedral trilayer graphene where Stoner phases were investigated at finite magnetic fields using compressibility measurements[11]. Possibly, also these phases show an insulating temperature and intriguing bias current dependence at $B$ = 0 T.

**Methods**

Sample fabrication and transport measurements

Bilayer graphene, graphite and hexagonal boron nitride (hBN) flakes were prepared by mechanical exfoliation of bulk crystals and identified via optical microscopy. The bilayer graphene flakes were then encapsulated in hBN and equipped with graphite top and bottom gates using a stamping technique with a polycarbonate (PC) film on top of a polydimethylsiloxane (PDMS) stamp[24,25]. In a first step, the heterostructures were assembled without top gate which was then added in a second stamping step. The device described in the main text was additionally equipped with graphite contacts and also used for measurements shown in Ref.[1] were this device is noted as device A and is described in detail. The sample shown in Extended Data Fig. 1 was fabricated in the same way but without graphite contacts. For this device, contacts were defined using electron-beam lithography and etched using an $SF_6$ plasma. The one-dimensional electrical contacts were then deposited by electron-beam evaporation of Cr/Au.

All electrical measurements (unless stated otherwise) were conducted in a dilution refrigerator equipped with home-made low-pass filters at a base temperature of 10 mK. The conductance was measured in a configuration shown in Fig. 1b/ Extended Data Fig. 1a using an a.c. bias current of 1 nA and an additional d.c. current where noted. Applying top gate voltages ($V_t$) and bottom gate voltages ($V_b$) allowed us to tune $n$ and $D$ as follows:

$$n = \frac{\varepsilon_0 \varepsilon_b}{d_b e} \left( \alpha V_t + V_b \right)$$

$$D = \frac{\varepsilon_b}{2 d_b} \left( \alpha V_t - V_b \right)$$

where $\varepsilon_0$ is the permittivity in vacuum, $\varepsilon_b$ is the dielectric constant of hBN, $d_b$ is the thickness of the bottom hBN flake, $e$ is the charge of an electron and $\alpha$ is the ratio of $d_b$ and the thickness of the top hBN flake. As contact resistance in graphite contacts depends on $T$, $I$, $n$, $D$ and $B_\perp$[1] we did not subtract contact resistance but mostly plotted $G$ in a.u.. Measured resistances $R = 1/G$ (with no subtracted contact resistances) are shown in Extended Data Fig. 2.

**Data availability**

The data that support the findings of this study are available from the corresponding authors upon reasonable request.

**Acknowledgements**

We thank Fan Zhang for fruitful discussions. R.T.W. and A.M.S. acknowledge funding from the Deutsche Forschungsgemeinschaft (DFG, German Research Foundation) under the SFB 1073 project B10. R.T.W. acknowledges partial funding from the SPP2244 from the Deutsche Forschungsgemeinschaft (DFG, German Research Foundataion). K.W. and T.T. acknowledge support from the JSPS KAKENHI (Grant Numbers 20H00354, 21H05233 and 23H02052) and World Premier International Research Center Initiative (WPI), MEXT, Japan.

**Author contributions**

A.M.S. fabricated the devices and conducted the measurements and data analysis. K.W. and T.T. grew the hexagonal boron nitride crystals. N.J, Z.D and L.S.L contributed the theoretical part. All authors discussed and interpreted the data. R.T.W. supervised the experiments and the analysis. The manuscript was prepared by A.M.S., Z.D, L.S.L. and R.T.W. with input from all authors.

**Corresponding authors**

R. Thomas Weitz (thomas.weitz@uni-goettingen.de)

**Competing interests**

Authors declare no competing interests.

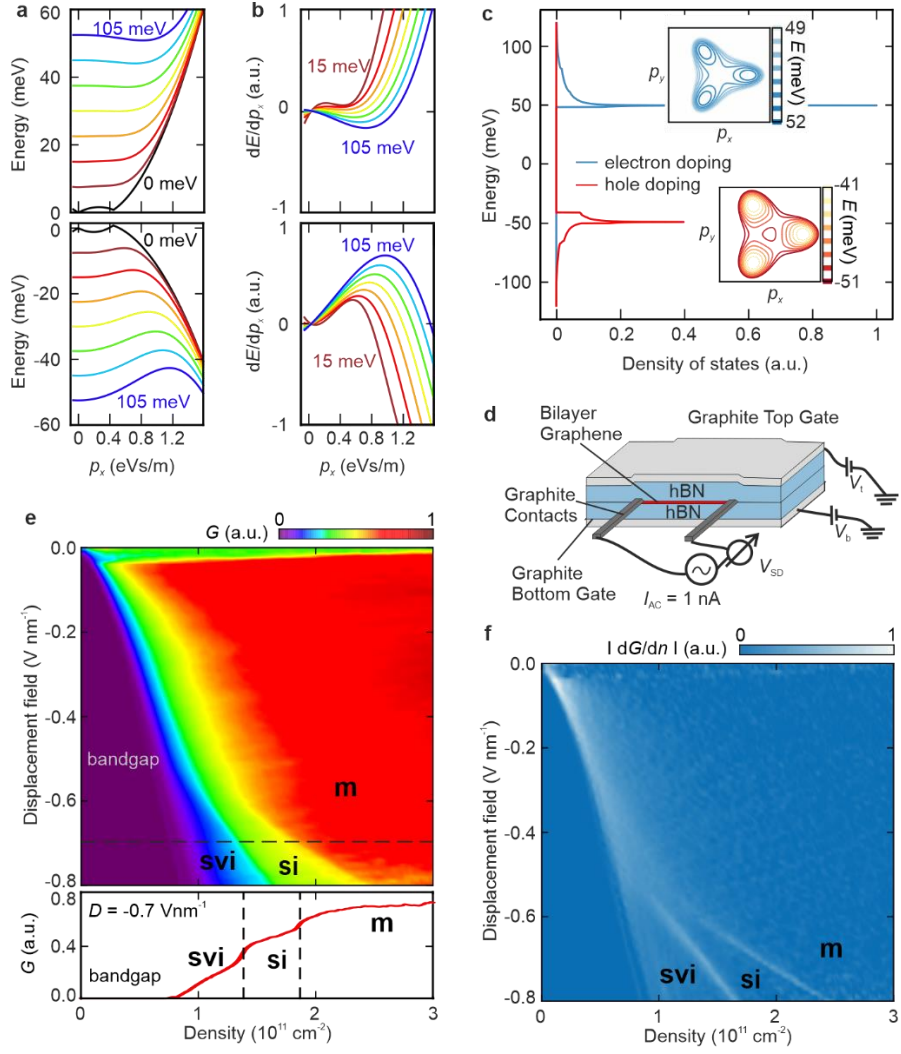

**Fig. 1. Density of states and phase transitions in electron-doped Bernal bilayer graphene without magnetic field. (a)** Calculated band structure of bilayer graphene at interlayer potential differences *U* ranging from 0 to 105 meV. Coupling parameters were taken from Ref.[26]. **(b)** Derivative of the energy bands with respect to the momentum in x-direction p$_x$. At large *U*, trigonal warping leads to a large |d*E*/d*p$_x$*| in the valence band while the conduction band is relatively flat (|d*E*/d*p$_x$*| $\approx$ 0). **(c)** Calculated density of states (DOS) as a function of the Fermi energy level at an interlayer potential difference *U* = 100 meV. Insets: Fermi surface contour at different Fermi energy levels at an interlayer potential difference *U* = 100 meV (please note the different energy scales in the valence and conduction band). At electron doping, trigonal warping is less pronounced and bands are flatter, i.e. changes in the Fermi surface topology appear in a smaller energy regime, resulting in a larger density of states at the band edge. **(d)** Schematic of the device as well as its electrical wiring used to conduct the transport measurement shown in the main text. **(e,f)** Conductance **(e)** and normalized derivative of the conductance **(f)** as a function of *n* and *D* with $B_\perp$ = $B_{\parallel}$ = 0 T. Steps in the conductance appear as peaks in the normalized derivative of the conductance. The **svi**, **si** and **m** phases are labeled.

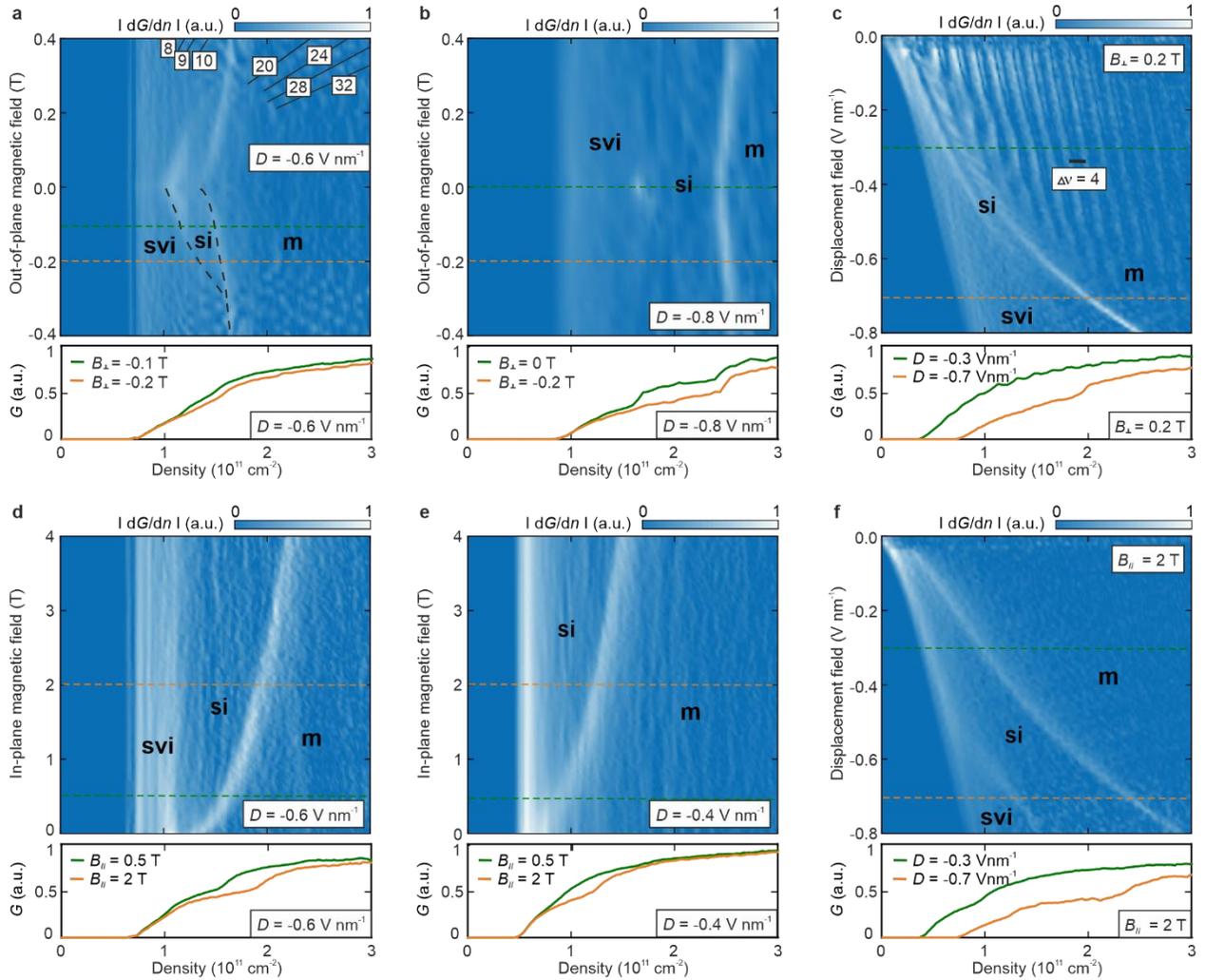

**Fig. 2. Magnetic field dependence of spin and valley ordered phases. (a,b)** Normalized derivative of the conductance as a function of $n$ and $B_\perp$ for $D$ = -0.6 V nm$^{-1}$ **(a)** and $D$ = -0.8 V nm$^{-1}$ **(b)**. Quantum Hall states are traced by lines and labeled by numerals for $B_\perp > 0$ T. At $D$ = -0.6 V nm$^{-1}$ they are one-fold degenerate in the density regime of the **svi** phase that becomes a spin and valley polarized Stoner metal at $B_\perp > 0.3$ T and four-fold degenerate in the density regime of the **m** phase. No quantum Hall states appear at $D$ = -0.8 V nm$^{-1}$. **(c)** Normalized derivative of the conductance as a function of $n$ and $D$ for $B_\perp$ = 0.2 T. Four-fold degenerate quantum Hall states appear in the density regime of the **m** phase. **(d,e)** Normalized derivative of the conductance as a function of $n$ and $B_{II}$ for $D$ = -0.6 V nm$^{-1}$ **(d)** and $D$ = -0.4 V nm$^{-1}$ **(e)**. **(f)** Normalized derivative of the conductance as a function of $n$ and $D$ for $B_{II}$ = 2 T. Line traces of the conductance $G$ as a function of $n$ are shown below each plot for different values of $D$ and $B$.

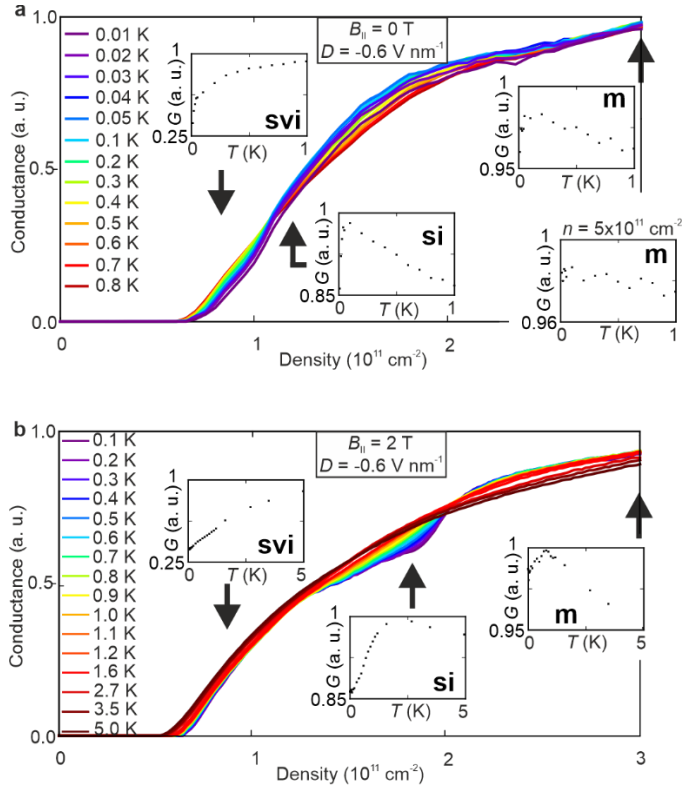

**Fig. 3. Temperature dependence of spin and valley ordered phases. (a,b)** Conductance $G$ as a function of the charge carrier density for $D$ = -0. 6 V nm$^{-1}$ and $B_{\parallel}$ = 0 T **(a)** and $B_{\parallel}$ = 2 T **(b)** for different temperatures. (insets) $G$ as a function of temperature $T$. Please note the different temperature scales in the insets of **a)** and **b)** )The corresponding densities are marked by an arrow ((**a**) **svi** phase: $n$ = 0.8 x 10$^{11}$ cm$^{-2}$, **si** phase: $n$ = 1.2 x 10$^{11}$ cm$^{-2}$, **m** phase: $n$ = 3 x 10$^{11}$ cm$^{-2}$ and $n$ = 5 x 10$^{11}$ cm$^{-2}$; **(b)** **svi** phase: $n$ = 0.8 x 10$^{11}$ cm$^{-2}$, **si** phase: $n$ = 1.8 x 10$^{11}$ cm$^{-2}$, **m** phase: $n$ = 3 x 10$^{11}$ cm$^{-2}$).

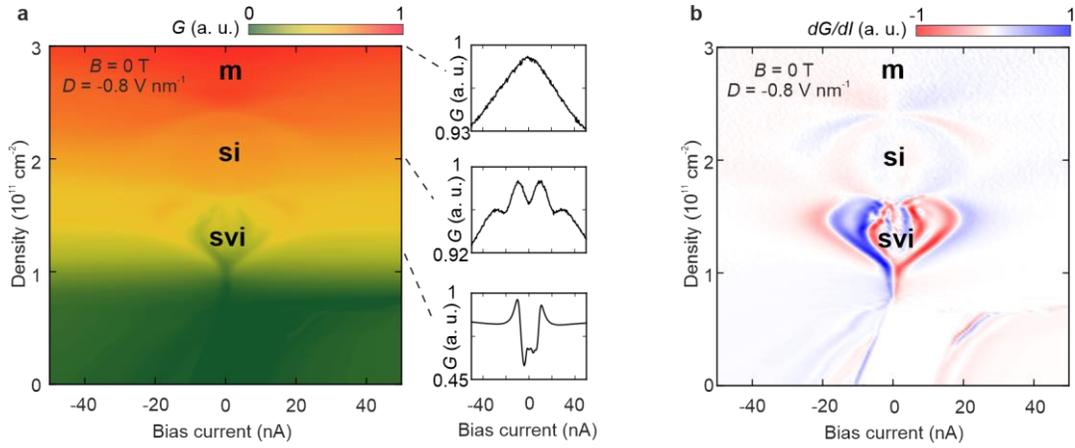

**Fig. 4. Bias current dependence of spin and valley ordered phases. (a)** Conductance in arbitrary units (a.u.) as a function of the bias current and the charge carrier density at $D$ = -0. 8 V nm$^{-1}$ and $B$ = 0 T. Line traces of the conductance as a function of the bias current are shown in the right panel for the **svi** phase ($n$ = 1.2 x 10$^{11}$ cm$^{-2}$), the **si** phase ($n$ = 2.0 x 10$^{11}$ cm$^{-2}$) and the **m** phase ($n$ = 3.0 x 10$^{11}$ cm$^{-2}$). **(b)** Derivative of the conductance with respect to bias current ($I$) as a function of $I$ and the charge carrier density at $D$ = -0. 8 V nm$^{-1}$ and $B$ = 0 T.

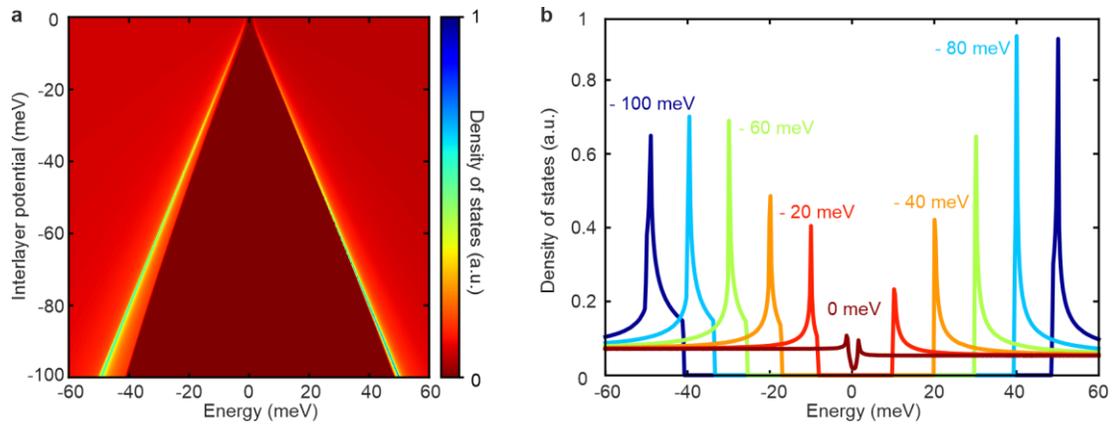

**Extended Data Fig. 1. Density of states in bilayer graphene. (a)** Calculated density of states as a function of the Fermi energy and the interlayer potential difference. Coupling parameters were taken from Ref.[26]. **(b)** Calculated density of states as a function of the Fermi energy for different interlayer potential differences. Due to electron-hole asymmetry, the peak in the density of states in electron doped bilayer graphene is located closer to the band edge than the peak in hole doped bilayer graphene and is more pronounced at large interlayer potential differences.

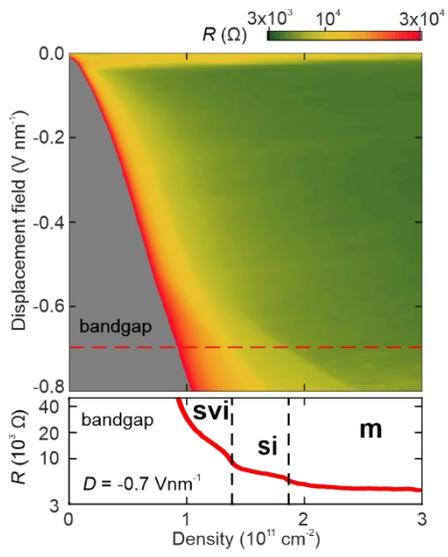

**Extended Data Fig. 2. Measured resistance in the svi, si and m phases**. Resistance as a function of $n$ and $D$ with $B_\perp = B_{II} = 0$ and line cut showing the resistance as a function of $n$ at $D$ = -0.7 Vnm$^{-1}$. The **svi**, **si** and **m** phases are labeled.

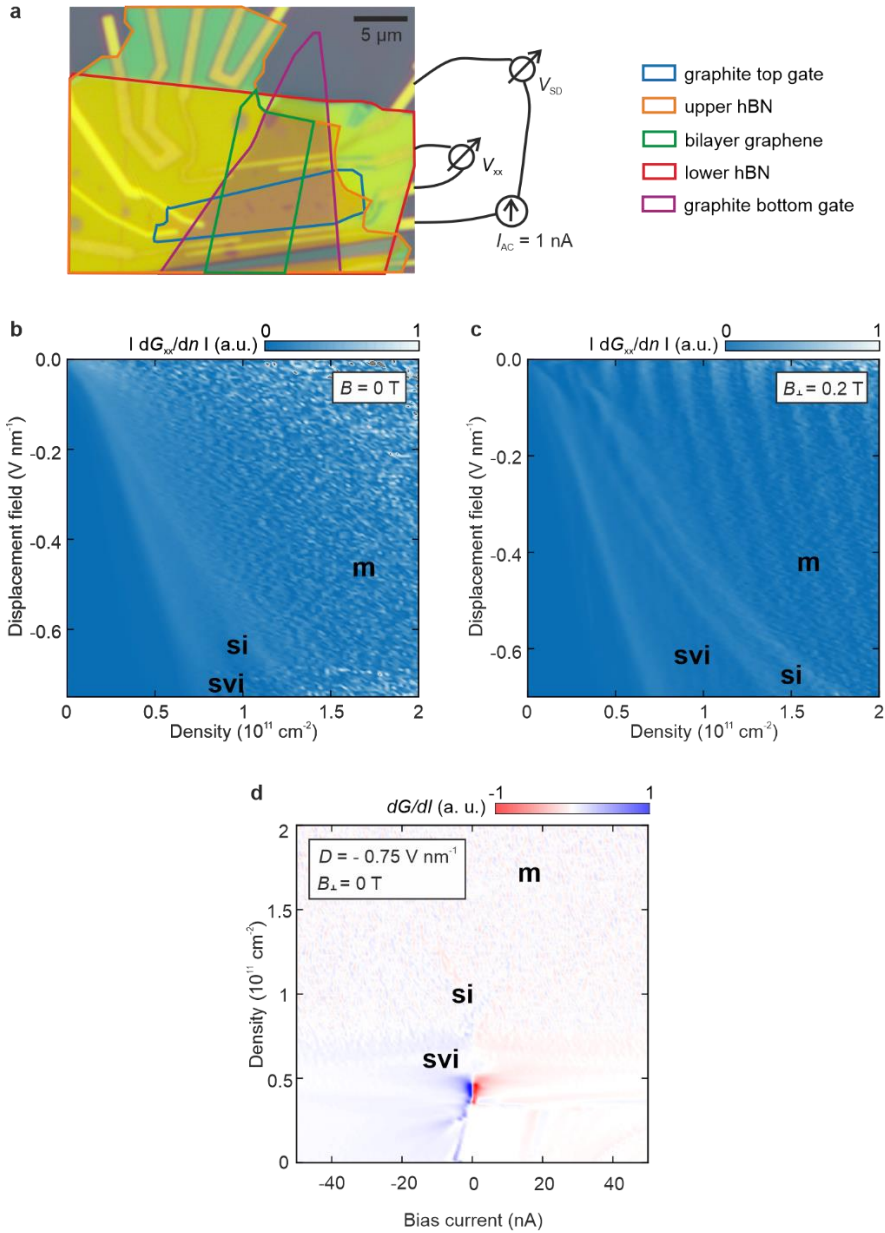

**Extended Data Fig. 3. Measurements conducted in a second device. (a)** Optical image of a second device. The corresponding hBN thicknesses were 69 nm (bottom hBN) and 30 nm (top hBN). **(b,c)** Normalized derivative of the conductance as a function of $n$ and $D$ with $B_\perp = B_{II} = 0$ T (b) and $B_\perp = 0.2$ T, $B_{II} = 0$ T (c) measured in the device shown in (a). The three different phases are labeled and show similar behavior as the device analyzed in the main text. **(d)** Derivative of the conductance with respect to $I$ as a function of $I$ and $n$ at $B_\perp = B_{II} = 0$ T and $D = -0.75$ V nm$^{-1}$ measured in the same device. While a gap is clearly present in the **svi** phase, a gap in the **si** phase is only weakly visible.

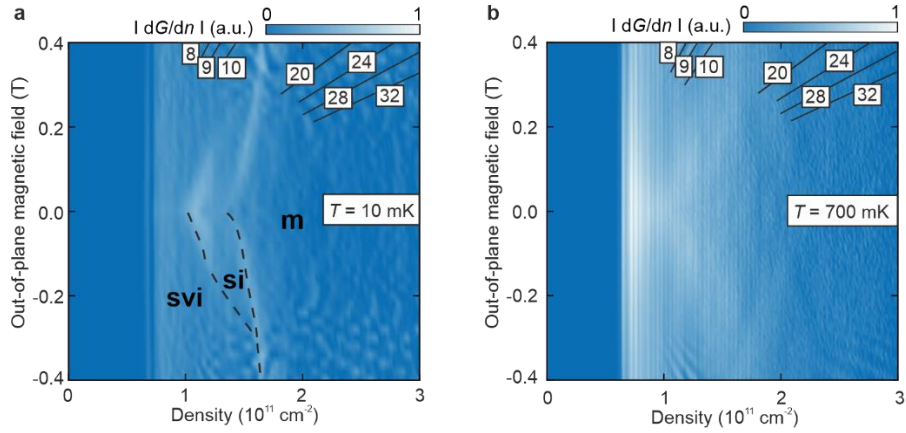

**Extended Data Fig. 4. Quantum Hall states at $D$ = -0.6 V nm$^{-1}$. (a,b)** Normalized derivative of the conductance plotted as a function of charge carrier density and out-of-plane magnetic field at a temperature of 10 mK **(a)** and 700 mK **(b)**. Quantum Hall states are traced by lines for positive out-of-plane magnetic fields. Corresponding filling factors are labeled by numerals. In the density regime of the **svi** phase, Landau levels are one-fold degenerate due to spin and valley polarization, in the density regime of the **m** phase they are four-fold degenerate due to spin and valley degeneracy. The one-fold degenerate quantum Hall states are even present at 700 mK when the **svi** and **si** phases at low $B$ are not stable anymore.

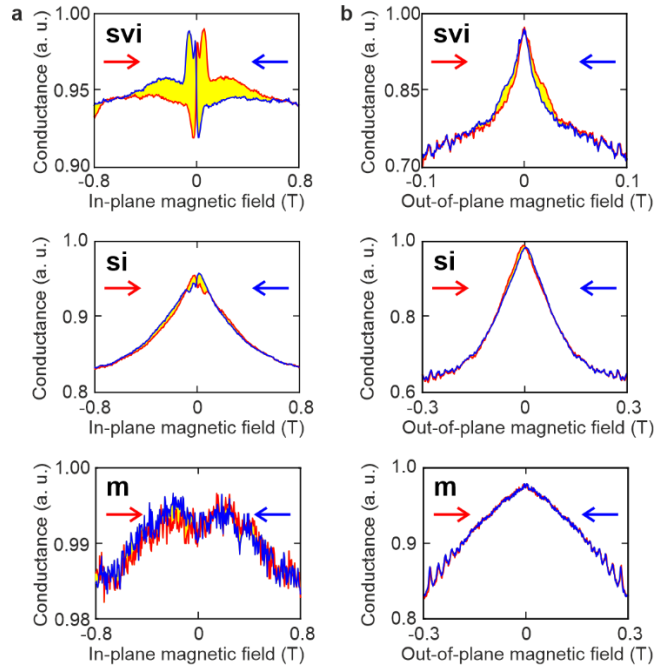

**Extended Data Fig. 5. Magnetic hysteresis of spin and valley ordered phases. (a,b)** In-plane magnetic hysteresis **(a)** and out-of-plane magnetic hysteresis **(b)** of the **svi**, **si** and **m** phases at $D$ = -0.6 V nm$^{-1}$. The corresponding charge carrier densities are $n$ = 1.0 x 10$^{11}$ cm$^{-2}$ (**svi** phase), n = 1.2 x 10$^{11}$ cm$^{-2}$ (**si** phase) and n = 3.0 x 10$^{11}$ cm$^{-2}$ (**m** phase). The forward sweeps are shown in red and the backward sweeps in blue. The hysteresis loop areas are shaded in yellow. Quantum oscillations appear at finite out-of-plane magnetic fields in the **m** phase.

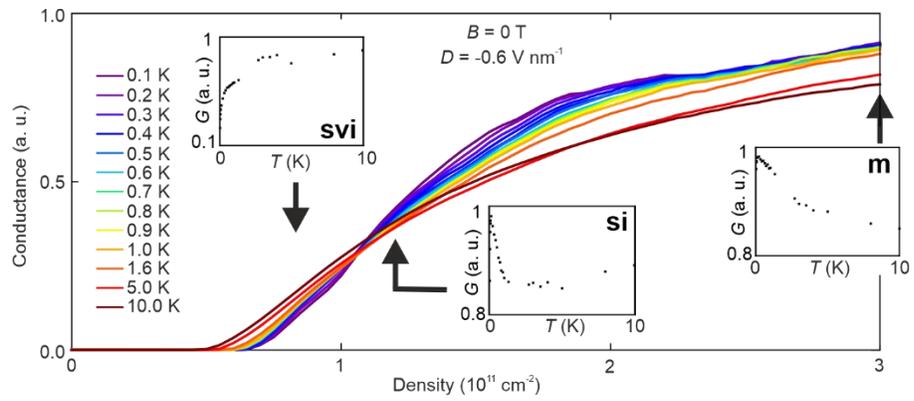

**Extended Data Fig. 6. Large temperature regime.** Conductance as a function of $n$ at different temperatures at $B$ = 0 T and $D$ = -0.6 Vnm$^{-1}$. The insets show linecuts of the conductance as a function of temperature for the different phases (si phase: $n$ = 0.8 x 10$^{11}$ cm$^{-2}$, svi phase: $n$ = 1.2 x 10$^{11}$ cm$^{-2}$, m phase: $n$ = 3 x 10$^{11}$ cm$^{-2}$). The **svi** phase is insulating until T ≥ 10 K.

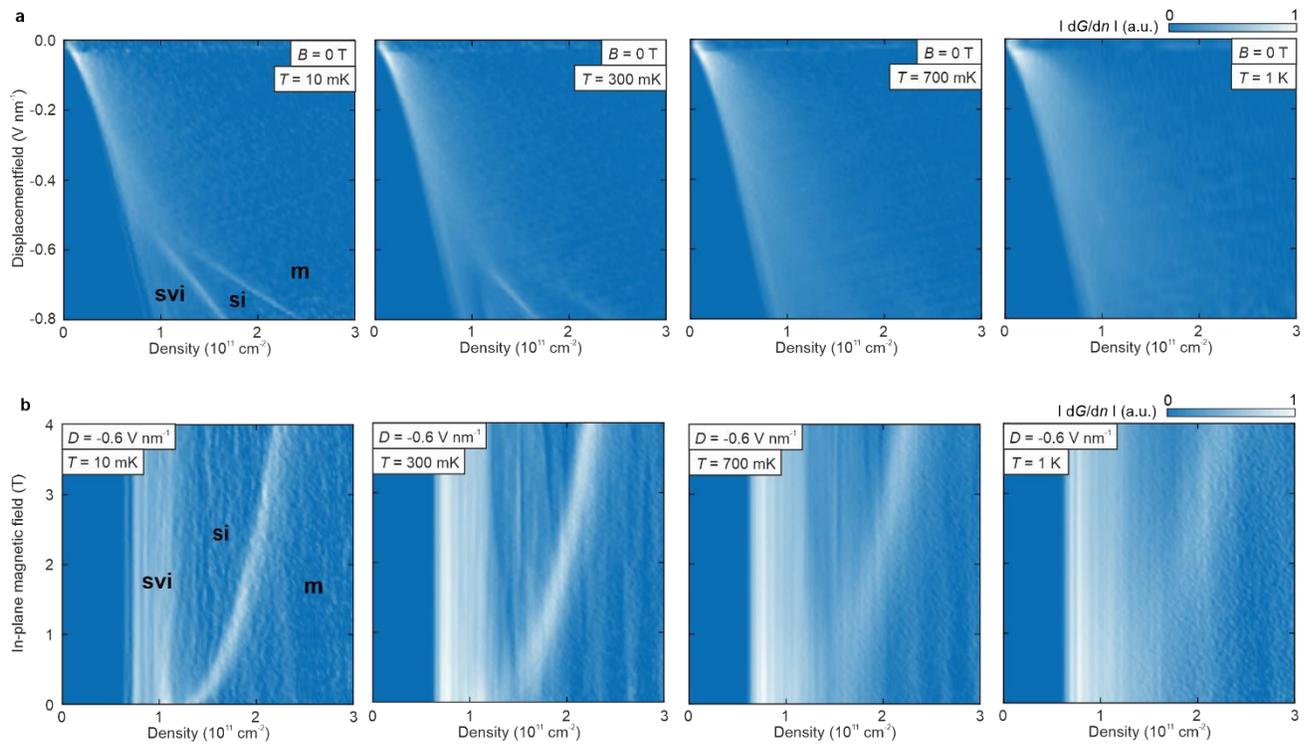

**Extended Data Fig. 7. Magnetic and electric field dependence of spin and valley ordered phases at large temperatures. (a)** Normalized derivative of the conductance as a function of charge carrier density and displacement field at zero magnetic field and at temperatures of 10 mK, 300 mK, 700 mK and 1 K. **(b)** Normalized derivative of the conductance as a function of charge carrier density and in-plane magnetic field at a displacement field of -0.6 V nm$^{-1}$ and at temperatures of 10 mK, 300 mK, 700 mK and 1 K.

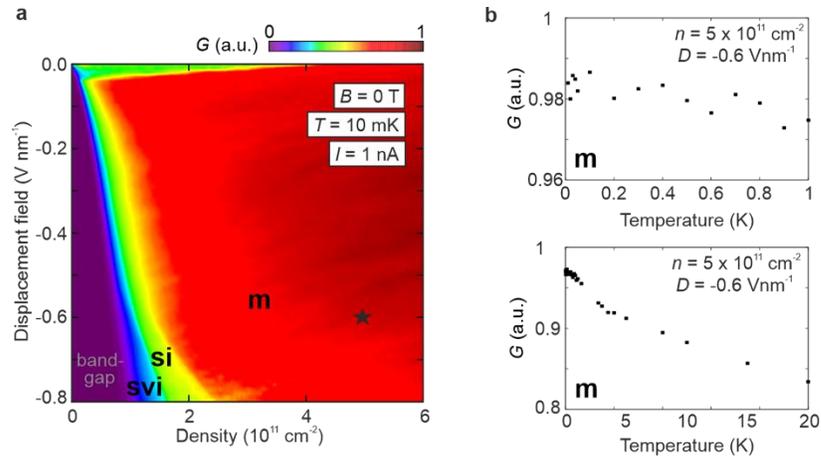

**Extended Data Fig. 8. Large density regime of the m phase. (a)** Conductance as a function of *n* and *D* with $B_\perp = B_{II}$ = 0 T. **(b)** Line traces of the conductance as a function of temperature at *D* = -0.6 Vnm⁻¹ and *n* = 5 x 10¹¹cm⁻² (metallic phase, marked with a star in (a)).

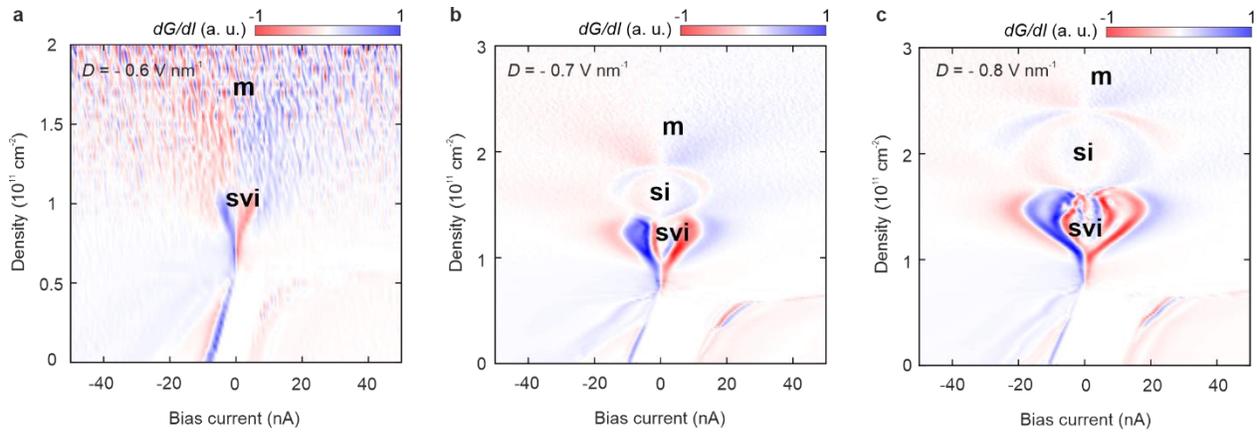

**Extended Data Fig. 9. Bias current dependence for different displacement fields. (a-c)** Derivative of the conductance as a function of bias current and charge carrier density at $B$ = 0 T, $T$ = 10 mK and $D$ = -0.6 V nm$^{-1}$ **(a)**, $D$ = -0.7 V nm$^{-1}$ **(b)** and $D$ = -0.8 V nm$^{-1}$ **(c)**. The different phases are labeled.

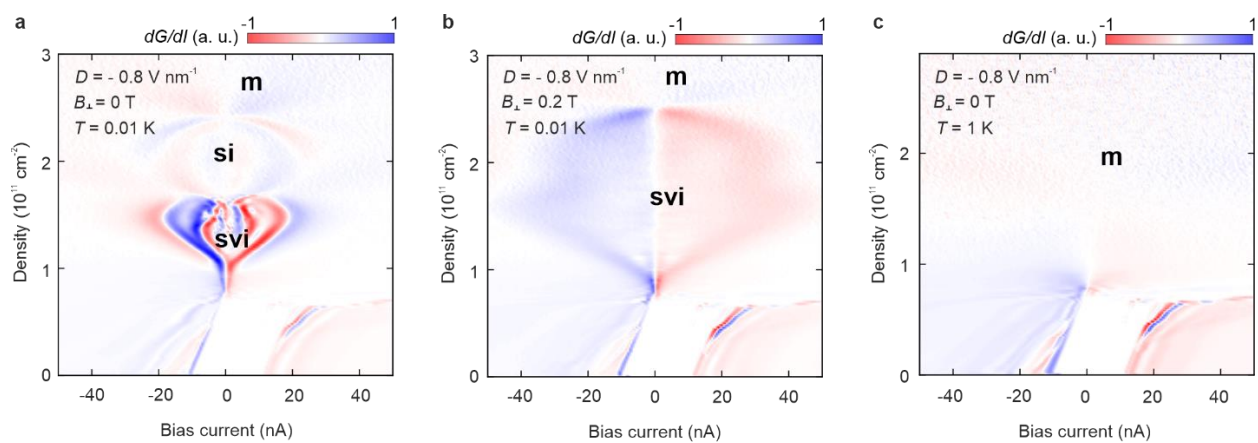

**Extended Data Fig. 10. Bias current dependence at *D* = -0.8 V nm⁻¹. (a-c)** Derivative of the conductance as a function of bias current and charge carrier density at *D* = -0.8 V nm⁻¹ and *B* = 0 T and *T* = 10 mK **(a)**, *B*⊥ = 0.2 T and *T* = 10 mK **(b)** and *B* = 0 T and *T* = 1 K **(c)**.

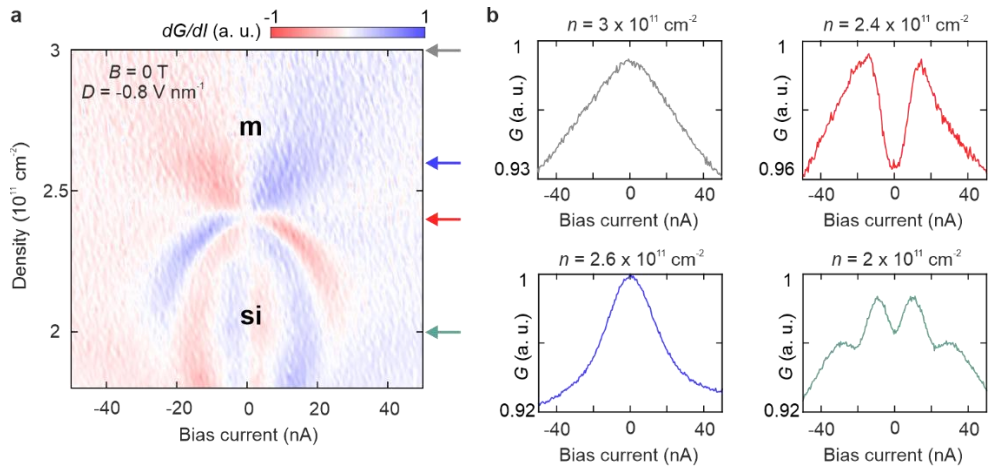

**Extended Data Fig. 11. Bias current dependence near the phase boundary between the si and m phases.**
**(a)** Derivative of the conductance with respect to bias current ($I$) as a function of $I$ and the charge carrier density at $D$ = -0. 8 V nm$^{-1}$ and $B$ = 0 T. **(b)** Line traces of the conductance as a function of $I$ at different charge carrier densities that are marked with arrows in **(a)**. The **m** phase exhibits a non-linear bias current dependence in the vicinity of the phase boundary between the **si** and **m** phase that becomes weaker with increasing charge carrier density, i.e. the peak becomes less pronounced and |d$G$/d$I$| becomes smaller. At the phase boundary ($n$ ≈ 2.4 x 10$^{11}$ cm$^{-2}$) the dependence on the applied bias current is weak. Please note the different scales in the conductance.

|  | **svi** phase | **si** phase | **m** phase |
|---|---|---|---|
| Valley polarization | yes | no | no |
| Spin polarization | yes | yes | no |
| Hysteresis in $B_{II}$ | yes | yes | no |
| Hysteresis in $B_\perp$ | yes | no | no |
| Onset of quantum Hall oscillations | $B_\perp \sim 300$ mT | $B_\perp > 300$ mT | $B_\perp < 200$ mT |
| Insulating $T$-dependence? | yes | yes | no (only near phase boundary to **si** phase) |
| $T_{crit}$ at $D$ = -0.6 Vnm$^{-1}$ and $B_{II}$ = 0 T | $\geq 4$ K | < 100 mK | - |
| $T_{crit}$ at $D$ = -0.6 Vnm$^{-1}$ and $B_{II}$ = 2 T | $\geq 5$ K | $\approx 2.5$ K | - |
| Non-linear bias-dependence? | yes | yes | no (only near phase boundary to **si** phase) |

**Extended Data Table 1. Summary of our main findings**. The table summarizes the behavior of the **svi, si** and **m** phases that is most consistent with experimental measurements shown and discussed in the main text.